\documentclass[english]{article}
\usepackage[T1]{fontenc}
\usepackage[latin1]{inputenc}
\usepackage{geometry}
\usepackage{amsmath}
\usepackage{setspace}
\usepackage{amssymb}

\makeatletter
\newcommand{\lyxaddress}[1]{
\par {\raggedright #1
\vspace{1.4em}
\noindent\par}
}

\usepackage{babel}
\makeatother
\begin{document}

\title{Quaternion Analysis for Generalized Electromagnetic Fields of Dyons
in Isotropic Medium}

\author{Jivan Singh$^{\text{(1)}}$, P. S. Bisht$^{\text{(2)}}$ and O. P.
S. Negi$^{\text{(2)}}$}

\maketitle

\lyxaddress{\begin{center}$^{\text{(1)}}$Department of Physics\\
 Govt. P. G. College\\
Pithoragarh -(UA), INDIA\par\end{center}}

\lyxaddress{\begin{center}$^{\text{(2)}}$Department of Physics\\
Kumaun University\\
S. S. J. Campus\\
Almora-263601 (UA), INDIA\par\end{center}}

\lyxaddress{\begin{center}Email:-$^{\text{(1)}}$jgaria@indiatimes.com\\
 $^{\text{(2)}}$ps\_bisht123@rediffmail.com\\
$^{\text{(2)}}$ ops\_negi@yahoo.co.in.\par\end{center}}

\begin{abstract}
Quaternion analysis of time dependent Maxwell's equations in presence
of electric and magnetic charges has been developed and the solutions
for the classical problem of moving charges (electric and magnetic)
are obtained in unique, simple and consistent manner.
\end{abstract}

\section{Introduction}

~~~~It is believed that in spite of the recent potential importance
of magnetic monopoles\cite{key-1},\cite{key-2},\cite{key-3} and
dyons \cite{key-4} towards the quark confinement problem \cite{key-5}
of quantum choromodynamics, possible magnetic condensation of vacuum
\cite{key-6}, CP-violation \cite{key-7}, their role as catalyst
in proton decay \cite{key-8} and the current grand unified theories
\cite{key-9}, the formalism necessary to describe them has been clumsy
and manifestly non-covariant. Keeping in view the potential importance
of monopoles and the results of Witten \cite{key-7} that monopoles
are necessarily dyons, we \cite{key-10,key-11} have also constructed
a self-consistent co-variant theory of generalized electromagnetic
fields associated with dyons each carrying the generalized charge
as complex quantity with its real and imaginary part as electric and
magnetic constituents. On the other hand quaternions were invented
by Hamilton \cite{key-12} to extend the theory of Complex numbers
to three dimensions. Maxwell's Equations of electromagnetism were
rewritten in terms of quaternions \cite{key-13}. Finklestein et al
\cite{key-14} developed the quaternionic quantum mechanics and Adler
\cite{key-15} described the theory of the algebraic structure of
quantum choromodynamics for strong interactions. Various aspects of
quaternions are discussed by Morita \cite{key-16} towards the kinematical
structure of Poincare gauge theory and the left-right Weinberg-Salam
theory of quantum choromodynamics. We have also studied \cite{key-17}
the quaternionic formulation for generalized field equations of dyons
in unique, simpler and compact notations. Quaternion non-Abelian gauge
theory has also been consistently discussed \cite{key-18} to maintain
the structural symmetry between the theory of linear gravity and electromagnetism.
It is also shown that quaternion formalism characterises the Abelian
and non-Abelian gauge structures \cite{key-19} of dyons in terms
of real and imaginary constituents of quaternion basis elements. Altenatively,
Kravchenko \cite{key-20} and his coworkers have consistently analysed
the Maxwell's equations for time-dependent electromagnetic fields
in homogeneous (isotropic) and chiral medium. Extending this, we \cite{key-21}
have also derived the generalised Maxwell's-Dirac equation in the
homogenous (isotropic) medium. It has been shown that the field equations
of dyons also remains invariant under the duality transformations
in isotropic homogeneous medium and the equation of motion reproduces
the rotationally symmetric gauge invariant angular momentum of dyons.
In order to extend the theory of monopoles and dyons in isotropic
medium and consequently the relevance of quaternion formalism of dyons,
in the present paper we have undertaken the study of the quaternion
analysis of time dependent Maxwell's equations in presence of electric
and magnetic charges and the solution for the classical problem of
moving sources are obtained in unique, simpler and consistent manner.
Quaternion forms of potential current, field equation and equation
of motion are developed in compact manner and it is imphasized that
the quantum equations in terms of quaternions are invariant under
quaternion, Lorentz and duality transformations. It has also been
discussed that the quaternion analyticity of dyons in isotropic medium
reproduces the results of Kravchenko \cite{key-20} in the absence
of magnetic monopole and accordingly this theory can be described
symmetrically for pure monopole in the absence of electric charge
or vice versa.

\section{Generalized field equation of dyons in Homogeneous Medium}

Assuming the existence of magnetic monopoles,we may write the following
form \cite{key-22} of symmetric generalized Maxwell-Dirac differential
equations \cite{key-1,key-21} in free space in SI units $($$c=\hbar=1)$
as;

\begin{eqnarray}
\overrightarrow{\nabla}.\overrightarrow{D} & = & \rho_{e}\nonumber \\
\overrightarrow{\nabla}.\overrightarrow{B} & = & \mu_{0}\rho_{m}\nonumber \\
\overrightarrow{\nabla}\times\overrightarrow{E} & = & -\frac{\partial\overrightarrow{B}}{\partial t}-\frac{\overrightarrow{j_{m}}}{\epsilon_{0}}\nonumber \\
\overrightarrow{\nabla}\times\overrightarrow{H} & = & \frac{\partial\overrightarrow{D}}{\partial t}+\overrightarrow{j_{e}}\label{eq:1}\end{eqnarray}
where $\rho_{e}$ and $\rho_{m}$ are respectively the electric and
magnetic charge densities while $\overrightarrow{j_{e}}$ and $\overrightarrow{j_{m}}$
are the corresponding current densities, $\overrightarrow{D}$ is
electric induction vector, $\overrightarrow{E}$ is electric field,
$\overrightarrow{B}$ is magnetic induction vector and $\overrightarrow{H}$
is magnetic field. Here we asume the homogenous (isotropic) medium
with the following definitions \cite{key-15},

\begin{eqnarray}
\overrightarrow{D} & = & \epsilon\overrightarrow{E}\,\,\,\,(\epsilon=\epsilon_{0}\epsilon_{r})\label{eq:2}\end{eqnarray}
and

\begin{eqnarray}
\overrightarrow{B} & = & \mu\overrightarrow{H}\,\,\,\,(\mu=\mu_{0}\mu_{r})\label{eq:3}\end{eqnarray}
where $\epsilon_{o}$ the free space permitivity, $\mu_{0}$ is the
permeability of free space and $\epsilon_{r}$ and $\mu_{r}$ are
defined respectively as relative permitivity and permeability in electric
and magnetic fields. On using equations (\ref{eq:2}) and (\ref{eq:3}),
equation (\ref{eq:1},\ref{eq:2},\ref{eq:3},\ref{eq:4}) takes the
following differential form \cite{key-19},

\begin{eqnarray}
\overrightarrow{\nabla}.\overrightarrow{E} & = & \frac{\rho_{e}}{\epsilon}\nonumber \\
\overrightarrow{\nabla}.\overrightarrow{B} & = & \mu\rho_{m}\nonumber \\
\overrightarrow{\nabla}\times\overrightarrow{E} & = & -\frac{\partial\overrightarrow{B}}{\partial t}-\frac{\overrightarrow{j_{m}}}{\epsilon}\nonumber \\
\overrightarrow{\nabla}\times\overrightarrow{B} & = & \frac{1}{v^{2}}\frac{\partial\overrightarrow{E}}{\partial t}+\mu\overrightarrow{j_{e}}\label{eq:4}\end{eqnarray}
Differential equations (\ref{eq:1},\ref{eq:2},\ref{eq:3},\ref{eq:4})
are the generalised field equations of dyons in homogenous medium
and the electric and magnetic fields are corresponding called generalised
electromegnetic fields of dyons. These electric and magnetic fields
of dyons are expresed in following differential form in homogenous
medium in terms of two potentials \cite{key-11}as,

\begin{eqnarray}
\overrightarrow{E} & = & -\overrightarrow{\nabla}\phi_{e}-\frac{\partial\overrightarrow{C}}{\partial t}-\overrightarrow{\nabla}\times\overrightarrow{D}\label{eq:5}\\
\overrightarrow{B} & = & -\overrightarrow{\nabla}\phi_{m}-\frac{1}{v^{2}}\frac{\partial\overrightarrow{D}}{\partial t}+\overrightarrow{\nabla}\times\overrightarrow{C}\label{eq:6}\end{eqnarray}
where $\{ C^{\mu}\}=\{\phi_{e},v\overrightarrow{C}\}$ and $\{ D^{\mu}\}=\{ v\phi_{m},\overrightarrow{D}\}$
are the two four-potentials associated with electric and magnetic
charges. 

Let us define the complex vector field $\overrightarrow{\psi}$in
the following form

\begin{eqnarray}
\overrightarrow{\psi} & = & \overrightarrow{E}-iv\overrightarrow{B}\label{eq:7}\end{eqnarray}
Equations (\ref{eq:5},\ref{eq:6}) and (\ref{eq:7}) lead to the
following relation between generalized field and the components of
generalized four-potential as,

\begin{eqnarray}
\overrightarrow{\psi} & = & -\frac{\partial\overrightarrow{V}}{\partial t}-\overrightarrow{\nabla}\phi-iv\,(\overrightarrow{\nabla}\times\overrightarrow{V)}\label{eq:8}\end{eqnarray}
where $\{ V_{\mu}\}$ is the generalised four-potential of dyons in
homogenous medium and defined as

\begin{eqnarray}
V_{\mu} & = & \{\phi,\overrightarrow{V}\}\label{eq:9}\end{eqnarray}
i.e. \begin{eqnarray}
\phi & = & \phi_{e}-iv\phi_{m}\label{eq:10}\end{eqnarray}
 and 

\begin{eqnarray}
\overrightarrow{V} & = & \overrightarrow{C}-i\frac{\overrightarrow{D}}{v}.\label{eq:11}\end{eqnarray}
Maxwell's field equation (\ref{eq:7},\ref{eq:8},\ref{eq:9},\ref{eq:10})
may then be written in terms of generalized field $\overrightarrow{\psi}$as

\begin{eqnarray}
\overrightarrow{\nabla}\cdot\overrightarrow{\psi} & = & \frac{\rho}{\epsilon}\label{eq:12}\\
\overrightarrow{\nabla}\times\overrightarrow{\psi} & = & -iv(\mu\overrightarrow{j}+\frac{1}{v^{2}}\frac{\partial\overrightarrow{\psi}}{\partial t})\label{eq:13}\end{eqnarray}
where $\rho$ and $\overrightarrow{j}$ are the generalized charge
and current source densities of dyons in homogenous medium given by

\begin{eqnarray}
\rho & = & \rho_{e}-i\,\frac{\rho_{m}}{v}\label{eq:14}\\
\overrightarrow{j} & = & \overrightarrow{j_{e}}-i\, v\,\overrightarrow{j_{m}}.\label{eq:15}\end{eqnarray}
Using equation (\ref{eq:12}) we introduce a new parameter$\overrightarrow{S}$
(i.e. the field current) as

\begin{eqnarray}
\overrightarrow{S} & =\square\overrightarrow{\psi} & =-\mu\frac{\partial\overrightarrow{j}}{\partial t}-\frac{1}{\epsilon}\overrightarrow{\nabla}\rho-iv\mu(\overrightarrow{\nabla}\times\overrightarrow{j})\label{eq:16}\end{eqnarray}
where $\square$is the D'Alembertian operator and expressed as

\begin{eqnarray}
\square & =\frac{1}{v^{2}}\frac{\partial^{2}}{\partial t^{2}}-\nabla^{2} & =\frac{1}{v^{2}}\frac{\partial^{2}}{\partial t^{2}}-\frac{\partial^{2}}{\partial x^{2}}-\frac{\partial^{2}}{\partial y^{2}}-\frac{\partial^{2}}{\partial z^{2}}.\label{eq:17}\end{eqnarray}
where $v$ is the speed of electromagnetic wave in homogenous isotropic
medium. In terms of complex potential the field equation is written
as

\begin{eqnarray}
\square\phi & = & v\mu\rho\label{eq:18}\\
\square\overrightarrow{V} & = & \mu\overrightarrow{j}.\label{eq:19}\end{eqnarray}
We write the following tensorial form of generalized Maxwell's -Dirac
equations of dyons in homogenous medium as

\begin{eqnarray}
F_{\mu\nu,\nu} & = & j_{\mu}^{e}\label{eq:20}\\
F_{\mu\nu,\nu}^{d} & = & j_{\mu}^{m}.\label{eq:21}\end{eqnarray}
Defining generalized field tensor of dyon as

\begin{eqnarray}
G_{\mu\nu} & = & F_{\mu\nu}-ivF_{\mu\nu}^{d}.\label{eq:22}\end{eqnarray}
One can directly obtain the following generalized field equation of
dyon in homogenous (isotropic) medium i.e.

\begin{eqnarray}
G_{\mu\nu,\nu} & = & j_{\mu}\label{eq:23}\\
G_{\mu\nu,\nu}^{d} & = & 0.\label{eq:24}\end{eqnarray}
The Lorentz four-force equation of motion for dyons in homogenous
(isotropic) medium as

\begin{eqnarray}
f_{\mu} & =m_{0}\ddot{x_{\mu}} & =ReQ*(G_{\mu\nu}u^{\nu})\label{eq:25}\end{eqnarray}
where $Re$ donetes real part, $\ddot{x_{\mu}}$ is the four-acceleration
and $\left\{ u^{\nu}\right\} $is the four-velocity of the particle
and $Q$ is the generalized charge of dyon in isotropic medium.

\section{Quaternion analysis for generalised Maxwell's equation in homogenous
medium}

A quaternion is defined as 

\begin{eqnarray}
q & = & q_{0}e_{0}+q_{1}e_{1}+q_{2}e_{2}+q_{3}e_{3}\label{eq:26}\end{eqnarray}
where $q_{0},q_{1},q_{2},q_{3}$ are real numbers and called the components
of the quaternion $q$ and the quaternion units $e_{0},e_{1},e_{2},e_{3}$
satisfy the following multiplication rules;

\begin{eqnarray}
e_{0}^{2} & = & 1\nonumber \\
e_{j}e_{k} & = & -\delta_{jk}+\epsilon_{jkl}e_{l}\label{eq:27}\end{eqnarray}
where $\delta_{jk}$ and $\epsilon_{jkl}$ (j, k, l= 1,2,3 and $e_{0}=1$)
are respectively the Kronecker delta and three-index Levi-Civita symbol.
The sum of two quaternions p and q is defined by adding the corresponding
components,

\begin{equation}
p+q=(p_{0}+q_{0})e_{0}+(p_{1}+q_{1})e_{1}+(p{}_{2}+q_{2})e_{2}+(p_{3}+q_{3})e_{3}\label{eq:28}\end{equation}
and the multiplication of two quaternions is defined as ;

\begin{eqnarray}
p.q & = & (p_{0}q_{0}-\overrightarrow{p}.\overrightarrow{q})e_{0}+\underset{j}{\Sigma}(p_{j}q_{0}+p_{0}q_{j}+\epsilon_{jkl}p_{k}q_{l})e_{j}\label{eq:29}\end{eqnarray}
Thus each quaternion $q$ is the sum of a scalar $q_{0}$ and a vector
$\overrightarrow{q}$,

\begin{eqnarray}
q & = & scalar(q)+vector(q)=q_{0}+\overrightarrow{q}\label{eq:30}\end{eqnarray}
where $\overrightarrow{q}=q_{1}e_{1}+q_{2}e_{2}+q_{3}e_{3}.$ Thus
the product of two quaternions $p$ and $q$ is also written as 

\begin{eqnarray}
pq & = & p_{0}q_{0}-\overrightarrow{p}.\overrightarrow{q}+p_{0}\overrightarrow{q}+q_{0}\overrightarrow{p}+\overrightarrow{p}\times\overrightarrow{q}\label{eq:31}\end{eqnarray}
where the dot and cross indicate, respectively, the usual three dimensional
scalar and vector products. For any quaternion, there exists a quaternion
conjugate

\begin{eqnarray}
\overline{q} & = & q-q_{1}e_{1}-q_{2}e_{2}-q_{3}e_{3}=q_{0}-\overrightarrow{q}.\label{eq:32}\end{eqnarray}
Quaternion conjugate is an automorphism of ring of quaternion i.e.

\begin{eqnarray}
(\overline{pq}) & = & (\overline{q})(\overline{p}).\label{eq:33}\end{eqnarray}
The norm of a quaternion is given as

\begin{eqnarray}
q.\overline{q} & = & \overline{q}.q=q_{0}^{2}+q_{1}^{2}+q_{2}^{2}+q_{3}^{2}=|q|^{2}.\label{eq:34}\end{eqnarray}
The inverse of a quaternion $q$ is also a quaternion

\begin{eqnarray}
q^{-1} & = & \frac{\overline{q}}{|q|^{2}}.\label{eq:35}\end{eqnarray}
Let us define the quaternionic form of differential operator as\cite{key-19},

\begin{eqnarray}
\boxdot & = & (-\frac{i}{v}\partial_{t}+D)\label{eq:36}\end{eqnarray}
and

\begin{eqnarray}
\overline{\boxdot} & = & (-\frac{i}{v}\partial_{t}-D)\label{eq:37}\end{eqnarray}
where $D=e_{1}\partial_{1}+e_{2}\partial_{2}+e_{3}\partial_{3}.$
As such we can express the quantum equation associated with generalized
four-potential, four current, electric field and magnetic field in
terms of quaternionic analysis as 

\begin{eqnarray}
V & = & -i\frac{\phi}{v}e_{0}+V_{1}e_{1}+V_{2}e_{2}+V_{3}e_{3}\label{eq:38}\\
J & = & -i\rho ve_{0}+J_{1}e_{1}+J_{2}e_{2}+J_{3}e_{3}\label{eq:39}\\
E & = & E_{1}e_{1}+E_{2}e_{2}+E_{3}e_{3}\label{eq:40}\\
B & = & B_{1}e_{1}+B_{2}e_{2}+B_{3}e_{3}.\label{eq:41}\end{eqnarray}
Operating equation (\ref{eq:37}) to equation (\ref{eq:7}) and using
equation (\ref{eq:27} ), we get

\begin{eqnarray}
\overline{\boxdot}\psi & = & iv\mu J=i\sqrt{\frac{\mu}{\epsilon}}J.\label{eq:42}\end{eqnarray}
Similarly we may operate equation (\ref{eq:36}) to equation (\ref{eq:39})
and on using equation (\ref{eq:27}) we get

\begin{eqnarray}
\boxdot J & = & S.\label{eq:43}\end{eqnarray}
where we have used the following subsidiary condition 

\begin{eqnarray}
\overrightarrow{\nabla}.\overrightarrow{V}+\frac{1}{v^{2}}\frac{\partial\phi}{\partial t} & = & 0\label{eq:44}\end{eqnarray}
and

\begin{eqnarray}
\overrightarrow{\nabla}.\overrightarrow{J}+\frac{\partial\rho}{\partial t} & = & 0\label{eq:45}\end{eqnarray}
Equation (\ref{eq:44}) is known as Lorentz gauge condition while
equation (\ref{eq:45}) is referred as continuity equation. $\overrightarrow{\psi}$
and $\overrightarrow{S}$ are defined as quaternion valued vector
functions in the following manner

\begin{eqnarray}
\psi & = & -\psi_{t}-\frac{i}{v}(e_{1}\psi_{1}+e_{2}\psi_{2}+e_{3}\psi_{3}\label{eq:46}\\
S & = & -S_{t}-i\sqrt{\frac{\epsilon}{\mu}}(e_{1}S_{1}+e_{2}S_{2}+e_{3}S_{3}).\label{eq:47}\end{eqnarray}
Similarly, we may obtain the quaternion conjugate field equations
for dyons in homogenous (isotropic) medium as

\begin{eqnarray}
\overline{\boxdot}\overline{V} & = & \overline{\psi}\label{eq:48}\\
\overline{\boxdot}\overline{J} & = & \overline{S}\label{eq:49}\end{eqnarray}
where $\overline{V},\,\overline{J},\,\overline{\psi}$ and $\overline{S}$
are the quaternion conjugates defined as

\begin{eqnarray}
\overline{V} & = & -i\frac{\phi}{v}e_{0}-(V_{1}e_{1}+V_{2}e_{2}+V_{3}e_{3})\label{eq:50}\\
\overline{J} & = & -i\rho ve_{0}-(J_{1}e_{1}+J_{2}e_{2}+J_{3}e_{3})\label{eq:51}\\
\overline{\psi} & = & -\psi_{t}+\frac{i}{v}(e_{1}\psi_{1}+e_{2}\psi_{2}+e_{3}\psi_{3})\label{eq:52}\\
\overline{S} & = & -S_{t}+i\sqrt{\frac{\epsilon}{\mu}}(e_{1}S_{1}+e_{2}S_{2}+e_{3}S_{3}).\label{eq:53}\end{eqnarray}
Hence the quaternion forms of equation (\ref{eq:18},\ref{eq:19})
for generalized potential, equation (\ref{eq:23},\ref{eq:24}) for
generalized Maxwell's Dirac equation and Lorentz force equation (\ref{eq:25})
for dyons in homogeneous medium may be expressed in terms of the following
set of quaternion equations in simple, compact and consistent manner. 

\begin{eqnarray}
\boxdot\overline{\boxdot}V & = & J\label{eq:54}\\
{}[\boxdot,G\,] & = & J\label{eq:55}\\
{}[\boxdot,J\,] & = & 0\label{eq:56}\\
{}[u,G\,] & = & f\label{eq:57}\end{eqnarray}
where

\begin{eqnarray}
u & = & -iu_{0}e_{0}+u_{1}e_{1}+u_{2}e_{2}+u_{3}e_{3}\label{eq:58}\\
G & = & -iG_{0}e_{0}+G_{1}e_{1}+G_{2}e_{2}+G_{3}e_{3}.\label{eq:59}\end{eqnarray}
In equation (\ref{eq:57}), $u$, $G$ and $f$ are the quaternionic
forms of velocity, generalized field tensor and Lorentz force associated
with dyons in homogeneous medium. The elements of G in equation (\ref{eq:59})
are the following quaternion forms,

\begin{eqnarray}
G_{0} & = & G_{01}e_{1}+G_{02}e_{2}+G_{03}e_{3},\label{eq:60}\\
G_{1} & = & iG_{10}e_{1}+G_{12}e_{2}+G_{13}e_{3},\label{eq:61}\\
G_{2} & = & iG_{20}e_{1}+G_{21}e_{2}+G_{23}e_{3},\label{eq:62}\\
G_{3} & = & iG_{30}e_{1}+G_{31}e_{2}+G_{32}e_{2}.\label{eq:63}\end{eqnarray}
Equations (\ref{eq:55}, \ref{eq:56} ) and (\ref{eq:57}) may also
be described as

\begin{eqnarray}
[\boxdot,G_{\mu}] & = & J_{\mu}\label{eq:64}\end{eqnarray}
and

\begin{eqnarray}
[u,G_{\mu}] & = & f_{\mu}\label{eq:65}\end{eqnarray}
where $J_{\mu}$ and $f_{\mu}$ are the four current and four force
associated with generalized field of dyons in homogeneous (isotropic)
medium. Let us factorize the wave operator as the combination of quaternion
and its conjugate in the following manner\cite{key-20},

\begin{eqnarray}
(-\triangle-\frac{1}{v^{2}}\partial_{t}^{2}) & = & (-\frac{i}{v}\partial_{t}+D)(-\frac{i}{v}\partial_{t}-D).\label{eq:66}\end{eqnarray}
Moreover, each solution of the wave equation

\begin{eqnarray}
(\triangle+\frac{1}{v^{2}}\partial_{t}^{2})b & = & 0\label{eq:67}\end{eqnarray}
can be written in a simple and compact form as the sum of two functions
$A$ and $B$, which reduces to following set of differential equations 

\begin{eqnarray}
(D-\frac{i}{v}\partial_{t})A & = & j_{e}\label{eq:68}\\
(-D-\frac{i}{v}\partial_{t})B & = & j_{m}\label{eq:69}\end{eqnarray}
respectively for electric and magnetic charges of a particle described
as dyons.

\section{Moving dyons}

Let us define the generalized charge of dyons moving in generalized
electromagnetic field as ;

\begin{eqnarray}
Q & = & e-i\frac{g}{v}\label{eq:70}\end{eqnarray}
where $e$ is electronic charge and $g$ is magnetic charge. Let us
assume that a dyon is moving with a velocity $\overrightarrow{V}(t)$.
Thus, the electric charge density $\rho_{e}$ of a dyon may be expressed
as,

\begin{eqnarray}
\rho_{e}(t,x) & = & e\delta(x-S(t))\label{eq:71}\end{eqnarray}
where $S(t)$ is the trajectory of the electron and the current density
$\overrightarrow{j_{e}}$ is written as,

\begin{eqnarray}
\overrightarrow{j_{e}}(t,x) & = & \overrightarrow{V}(t)\rho_{e}(t,x).\label{eq:72}\end{eqnarray}
We now use equation (\ref{eq:66}) along with the known solution of
the equation i.e.

\begin{eqnarray}
(\triangle+\frac{1}{v^{2}}\partial_{t}^{2})b(t,x) & = & A(t)\delta(x-S(t))\label{eq:73}\end{eqnarray}
which may be expressed in terms of the following formula,

\begin{eqnarray}
b(t,x) & = & \frac{A(\tau(t))}{4\pi|x-S(\tau(t))|(1-M(S(t))}\label{eq:74}\end{eqnarray}
where

\begin{eqnarray}
M(\tau) & = & \frac{<\overrightarrow{V}(\tau),x-S(\tau)>}{v|x-S(\tau)|}\label{eq:75}\end{eqnarray}
is called the Mach number and $\tau$ satisfies the equation

\begin{eqnarray}
\frac{|x-S(\tau)|}{v}-(t-\tau) & = & 0.\label{eq:76}\end{eqnarray}
Similarly as the case of electron, let us describe that a monopole
constituent of a dyon is also moving with a velocity $\overrightarrow{V}(t)$.
Then the magnetic charge density $\rho_{m}$ and magnetic current
density $\overrightarrow{j_{m}}$ in view of duality transformations
\cite{key-21,key-22} lead to the following expressions i.e.

\begin{eqnarray}
\rho_{m}(t,x) & = & g\delta(x-S(t))\label{eq:77}\end{eqnarray}
and

\begin{eqnarray}
\overrightarrow{j_{m}}(t,x) & = & \overrightarrow{V}(t)\frac{\rho_{m}(t,x)}{v^{2}}.\label{eq:78}\end{eqnarray}
Taking into account the explicit form of $\rho_{e}$ , $j_{e}$ and
$\rho_{m}$, $j_{m}$ equation (\ref{eq:42}) is described as follows:

\begin{eqnarray}
\overline{\boxdot}\psi & = & [\frac{1}{\epsilon}+iv\mu\overrightarrow{V}(t)]e\delta(x-S(t))-\frac{i}{\epsilon}[\frac{1}{v}+i\frac{\overrightarrow{V}(t)}{v^{2}}]g\delta(x-S(t)).\label{eq:79}\end{eqnarray}
Thus, the purely vectorial biquaternion

\begin{eqnarray}
\psi(t,x) & = & (-\frac{i}{v}\partial_{t}+D)b(t,x)\label{eq:80}\end{eqnarray}
is a solution of equation (\ref{eq:79}) if $b$ is a solution of
equation (\ref{eq:73}) with

\begin{eqnarray}
A(t) & = & [(\frac{e}{\epsilon}+iev\mu\overrightarrow{V}(t))-\frac{i}{\epsilon}(\frac{g}{v}+ig\frac{\overrightarrow{V}(t)}{v^{2}})].\label{eq:81}\end{eqnarray}
Here $b$ and $A$ both are complex quaternionic functions. Using
equation (\ref{eq:73},\ref{eq:74},\ref{eq:75}) we obtain 

\begin{eqnarray}
b_{0}(t,x) & = & \frac{\frac{1}{\epsilon}(e-i\frac{g}{v})}{4\pi|x-S(\tau(t))|(1-M(\tau(t))}=\frac{Q}{4\pi\epsilon|x-S(\tau(t))|(1-M(\tau(t))}\label{eq:82}\end{eqnarray}
and

\begin{eqnarray}
\overrightarrow{b}(t,x) & = & \frac{i(ev\mu\overrightarrow{V}(t)-i\frac{g}{\epsilon}\frac{\overrightarrow{V}(t)}{v^{2}})}{4\pi|x-S(\tau(t))|(1-M(\tau(t))}=\frac{iv\mu\overrightarrow{V}(t)Q}{4\pi|x-S(\tau(t))|(1-M(\tau(t))}.\label{eq:83}\end{eqnarray}
Thus, solution (\ref{eq:79}) reduces to a simple differentiation,
i.e.

\begin{eqnarray}
\psi(t,x) & = & (-\frac{i}{v}\partial_{t}+D)[\frac{\frac{1}{\epsilon}Q+iv\mu\overrightarrow{V}(t)Q}{4\pi|x-S(\tau(t))|(1-M(\tau(t))}].\label{eq:84}\end{eqnarray}
We may now introduce the auxiliary functions as 

\begin{eqnarray}
\varphi & = & \frac{1}{|x-S(\tau(t))|(1-M(\tau(t))}\label{eq:85}\\
\zeta & = & \frac{1}{4\pi}[\frac{1}{\epsilon}Q+iv\mu\overrightarrow{V}(t)Q].\label{eq:86}\end{eqnarray}
Then

\begin{eqnarray}
\psi & = & (-\frac{i}{v}\partial_{t}+D)[\zeta].\varphi+(-\frac{i}{v}\partial_{t}+D)[\varphi].\zeta.\label{eq:87}\end{eqnarray}
It is easy to see that the scalar part of the expression on the right
hand side is zero. Rewriting Maxwell's equations (\ref{eq:4}) in
vector form as follows:

\begin{eqnarray}
\overrightarrow{\psi} & = & -\frac{i}{v}(\partial_{t}\overrightarrow{\zeta}.\varphi+\partial_{t}\varphi.\overrightarrow{\zeta)}+\varphi\overrightarrow{\nabla}\times\overrightarrow{\zeta}+\zeta_{0}\overrightarrow{\nabla}\varphi+[\overrightarrow{\nabla}\varphi\times\overrightarrow{\zeta}].\label{eq:88}\end{eqnarray}
By the definition of $\overrightarrow{\psi}$ from equation (\ref{eq:7}),
we have,

\begin{eqnarray}
\overrightarrow{E} & = & -\frac{i}{v}(\partial_{t}\overrightarrow{\zeta}.\varphi+\partial_{t}\varphi.\overrightarrow{\zeta})+\frac{1}{4\pi\epsilon}Q[\overrightarrow{\nabla}\varphi]\label{eq:89}\end{eqnarray}
and

\begin{eqnarray}
v\overrightarrow{B} & = & \varphi\overrightarrow{\nabla}\times\overrightarrow{\zeta}+[\overrightarrow{\nabla}\varphi\times\overrightarrow{\zeta}].\label{eq:90}\end{eqnarray}
To obtain the following generalized electric and magnetic field vectors
$\overrightarrow{E}$ and $\overrightarrow{B}$ in explicit form we
have used the following intermidate equalities i.e.

\begin{eqnarray}
\partial_{t}\overrightarrow{\zeta} & = & i\frac{(ev\mu\overrightarrow{V'}(\tau)-i\frac{g}{\epsilon}\overrightarrow{V'}(\tau))}{4\pi(1-M(\tau))},\label{eq:91}\end{eqnarray}

\begin{eqnarray}
\overrightarrow{\nabla}\times\overrightarrow{\zeta} & = & \frac{i}{4\pi}[ev\mu(\overrightarrow{\nabla}\tau\times\overrightarrow{V'}(\tau))-i\frac{g}{\epsilon}(\overrightarrow{\nabla}\tau\times\overrightarrow{V'}(\tau))],\label{eq:92}\end{eqnarray}

\begin{eqnarray*}
[\overrightarrow{\nabla}\varphi\times\overrightarrow{\zeta}] & = & (v^{2}+\frac{i}{4\pi}\{\frac{ev\mu[(x-s)\times\overrightarrow{V}]-i\frac{g}{\epsilon}[(x-s)\times\overrightarrow{V}]}{v^{2}|x-s|^{3}(1-M)^{3}}\}\times\end{eqnarray*}

\begin{eqnarray}
<v^{2}+<\overrightarrow{V}\,',x-s> & -|\overrightarrow{V}|^{2} & ).\label{eq:93}\end{eqnarray}
As such, the solutions of the above problem for a moving dyon in terms
of electric and magnetic field vectors of may now be obtained in the
following forms in terms of intermidate equalities i.e.,

\[
\overrightarrow{E}=\frac{i}{4\pi}e\mu\{\frac{\overrightarrow{V'}(\tau)}{|x-s|(1-M)^{2}}+[\frac{(\overrightarrow{V}(\tau)|x-s|-\overrightarrow{v}|x-s|)}{v|x-s|^{3}(1-M)^{3}}](v^{2}+<\overrightarrow{V'},x-s>-|\overrightarrow{V}|^{2})\}\]

\begin{equation}
+\frac{g}{4\pi\epsilon v}\{\frac{\overrightarrow{V'}(\tau)}{|x-s|(1-M)^{2}}+[\frac{(\overrightarrow{V'}(\tau)|x-s|-\overrightarrow{v}|x-s|)}{v|x-s|^{3}(1-M)^{3}}](v^{2}+<\overrightarrow{V'},x-s>-|\overrightarrow{V}|^{2})\}\label{eq:94}\end{equation}

and

\[
\overrightarrow{B}=\frac{e\mu}{4\pi}\{\frac{[(x-s)\times\overrightarrow{V'}(\tau)]}{v|x-s|^{2}(1-M)^{2}}+\frac{[(x-s)\times\overrightarrow{v}(\tau)]}{v^{2}|x-s|^{3}(1-M)^{3}}(v^{2}+<\overrightarrow{V'},x-s>-|\overrightarrow{V}|^{2})\}\]

\begin{equation}
-\frac{ig}{v\epsilon}\{\frac{[(x-s)\times\overrightarrow{V'}(\tau)]}{v|x-s|^{2}(1-M)^{2}}+\frac{[(x-s)\times\overrightarrow{v}(\tau)]}{v^{2}|x-s|^{3}(1-M)^{3}}(v^{2}+<\overrightarrow{V'},x-s>-|\overrightarrow{V}|^{2})\}.\label{eq:95}\end{equation}
These expressions, associated with the solutions of generalized field
equations of dyons, reduce to the solutions of usual elecrtic and
magnetic field vectors in the absence of magnetic (electric) charge
on dyons similar to those described by Kravchenko \cite{key-20} or
vice versa.

\end{document}